\documentclass{pasj01}
\Received{$\langle$reception date$\rangle$}
\Accepted{$\langle$acception date$\rangle$}
\Published{$\langle$publication date$\rangle$}
\usepackage[dvips]{graphicx}

\begin{document}

\title{Parallax of a Mira variable R Ursae Majoris studied with astrometric VLBI}
\author{
Akiharu \textsc{Nakagawa}$^{1}$, 
Tomoharu \textsc{Kurayama}$^{2}$, 
Makoto \textsc{Matsui}$^{1}$, 
Toshihiro \textsc{Omodaka}$^{1}$, 
Mareki \textsc{Honma}$^{3}$, 
Katsunori M \textsc{Shibata}$^{3}$, 
Katsuhisa \textsc{Sato}$^{3}$, 
and 
Takaaki \textsc{Jike}$^{3}$
}%
\altaffiltext{}{
$^{1}$Graduate School of Science and Engineering, Kagoshima University, 1-21-35 Korimoto, Kagoshima-shi, Kagoshima 890-0065, Japan \\
$^{2}$Teikyo University of Science, 2525 Yatsusawa, Uenohara-shi, Yamanashi 409-0193, Japan \\
$^{3}$Mizusawa VLBI Observatory, National Astronomical Observatory of Japan, 2-12 Hoshi-ga-oka, Mizusawa-ku, Oshu-shi, Iwate 023-0861, Japan \\
}
\email{nakagawa@sci.kagoshima-u.ac.jp}

%\KeyWords{key word${}_1$ --- key word${}_2$ --- \dots --- key word${}_n$}

\maketitle

%%%%%%%%%% Abstract %%%%%%%%%%
\begin{abstract}
 We have measured an annual parallax of the Mira variable R~Ursae~Majoris (R~UMa) with the VLBI exploration for Radio Astronomy (VERA). 
From the monitoring VLBI observations spanning about two years, we detected H$_2$O maser spots in the LSR velocities ranges from 37 to 42 km\,s$^{-1}$. 
We derived an annual parallax of 1.97$\pm$0.05\,mas, and it gives a corresponding distance of 508$\pm$13\,pc. 
The VLBI maps revealed 72 maser spots distributed in $\sim$110 au area around an expected stellar position. 
Circumstellar kinematics of the maser spots were also revealed by subtracting a systemic motion in the Hipparcos catalog from proper motions of each maser spots derived from our VLBI observations.  
Infrared photometry is also conducted to measure a $K$ band apparent magnitude, and we obtained a mean magnitude of $m_K$ = 1.19$\pm$0.02\,mag. 
Using the trigonometric distance, the $m_K$ is converted to a $K$ band absolute magnitude of $M_K = -$7.34$\pm$0.06\,mag. 
This result gives a much more accurate absolute magnitude of R~UMa than previously provided. 
We solved a zero-point of $M_K - \log P$ relation for the Galactic Mira variables and obtained a relation of $M_K = -$3.52 $\log P$ + (1.09 $\pm$ 0.14). 
Other long period variables including red supergiants, whose distances were determined from astrometric VLBI, were also compiled to explore the different sequences of $M_K - \log P$ relation. 
\end{abstract} 

%●%%%%%%%%%% Introduction %%%%%%%%%% 
\section{Introduction} 
Mira variables and Long Period Variables (LPVs) are low- to intermediate-mass ($1- 8 M_{\odot}$) asymptotic giant branch (AGB) stars that pulsate with a period range of 100 $-$ 1000 days. 
They are surrounded by large and extended dust and molecular shells, and in sources with mass-loss rate higher than $\sim 10^{-7} M_{\odot}$yr$^{-1}$, we sometimes find maser emissions of H$_2$O, SiO, or OH \citep{gai14, hab03}. 
Because of their high mass-loss rate, they are also important source to study chemical composition of the universe. 
Another characteristic of the sources is concerned to their periodic variation. 
The relation between K band apparent magnitude ($m_K$) and logarithm of pulsation period ($\log P$) of Mira variables is well known in the Large Magellanic Cloud (LMC) \citep{fea89,ita04-1}. 
If the LMC distance is given, the relation can be converted to a relation of absolute magnitude ($M_K$) and $\log P$, then it can be used as a distance indicator. 
Since there is a metallicity difference between LMC and our galaxy, it is also important to establish this $M_K-\log P$ relation using sources in our own galaxy. 
Since the LPVs are very bright in infrared, we can use them to probe a region where interstellar extinction is strong such as the direction of the Galactic Center and Galactic plane. 
However, a construction of the $M_K-\log P$ relation for the Galactic Mira variables has long been difficult because of large errors in $M_K$ due to distance uncertainty. 
Making use of a high performance of the VERA array \citep{kob03}, which is a Japanese VLBI project dedicated to the Galactic astrometry, we aim to construct the $M_K-\log P$ relation for Galactic LPVs. 

R Ursae Majoris (R~UMa) is an O-rich Mira variable \citep{kna00} with a pulsation period of 301.6 days (GCVS)\footnote[1]{General Catalog of Variable Stars\\http://heasarc.gsfc.nasa.gov/W3Browse/all/gcvs.html}. 
The H$_2$O masers associated with R~UMa also exhibit regular periodic variation with a phase lag of $\sim$0.3 with respect to the optical light curve \citep{shi08}. 
Although the Hipparcos satellite has measured the annual parallax of 2.37$\pm$1.06 mas \citep{van07}, its corresponding distance of 422$^{+341}_{-130}$ pc still has a large error and it brings an uncertainty to the absolute magnitude. 
In this study, we observe this source with VERA to obtain more accurate distance based on an precise astrometry of H$_2$O maser positions. 

A high resolution VLBI map gives an angular distribution of maser spots around the star. 
Then, a time series of multiple VLBI observations can add kinematic information to the masers. 
This helps us to understand a global picture of circumstellar medium on the schemes of time and space. 
In case the maser distribution is relatively isotropic like a Mira variable `T~Lep' in our previous study \citep{nak14}, a circumstellar kinematics can be derived from an analysis only using VLBI maps. 
The detailed procedure is given in section 3.4 in \citet{nak14}. 
On the other hand, if the number of the maser spots is small or the distribution is far from isotropic, the same procedure can not offer a reliable kinematic picture of the maser spots. 
In this study, we propose a new method to obtain a kinematics of the circumstellar masers in R~UMa. 

We solve the $M_K-\log P$ relation of the Galactic Mira variables based on the latest results from astrometric VLBI observations. 
Infrared photometry is also used to measure an apparent magnitude in $K$ band.

%●%%%%%%%%% Observation and Data Reduction %%%%%%%%%%
\section{Observation and Data Reduction}
\subsection{VLBI Observations}
\label{sec_obs}
We have observed H$_2$O maser emission at the rest frequency of 22.235080\,GHz (6$_{16}$-5$_{23}$ transition) associated with R~UMa at 24 epochs between 2006 April and 2008 July with a typical separation of 1--2 months using the VLBI Exploration of Radio Astrometry (VERA). 
The VERA consists of four antennas of 20\,m aperture at Mizusawa, Iriki, Ogasawara, and Ishigaki-jima \citep{kob03}. 
An extra-galactic continuum source J1056$+$7011 was also observed simultaneously with a dual-beam system equipped with the antennas \citep{kaw00} to maximize on-source integration time of each source. 
This continuum source is used as a position reference in our phase-referencing analysis. 
Coordinates of the two sources in J2000 equinox are shown in table\ref{coordinate}.
The two sources are separated by 1.78$^{\circ}$ with a position angle of 39$^{\circ}$ on the sky plane. 
We summarize the observations in table~\ref{obs_table}. 
Observation dates and system noise temperatures ($T_{\mathrm{sys}}$) in each station are presented. 

%●データ記録と相関処理について %
The received source signals from two receivers were recorded using the SONY DIR\,2000 system with a recording rate of 1024 Mbps in 2-bit quantization, yielding a data with its total bandwidth of 256 MHz. 
This total bandwidth of 256 MHz was divided into 16 IF channels of 16 MHz bandwidth. 
One IF channel was assigned for the maser emission from R~UMa, while the other 15 IF channels were assigned for continuum emission from the reference source J1056$+$7011. 
Correlation was made with the Mitaka FX correlator \citep{shi98} using the coordinates in table~\ref{coordinate}. 
In the correlation process of R~UMa, an 8\,MHz data including H$_2$O maser lines out of the 16 MHz data was selected and divided into 512 spectral channels to obtain a frequency spacing of 15.625 kHz, which corresponds to a velocity spacing of 0.2108 km\,s$^{-1}$. 
For the last observation (Obs. ID 24), a broader frequency spacing of 31.25 kHz corresponding to a velocity spacing of 0.4216 km\,s$^{-1}$ was adopted. 
For the data of reference source J1056$+$7011, each IF channel data were divided into 64 spectral channels.
The synthesized beam size was typically 1.2\,mas$\times$0.8\,mas with its major axis position angle of 155$^{\circ}$. 

%●データ解析について %
Calibration and imaging were performed using the Astronomical Image Processing System (AIPS) software package. 
We performed an amplitude calibration using $T_{\mathrm{sys}}$. 
Bright continuum sources 3C273B and 4C39.25 were used in a bandpass calibration of the receiver of R~UMa, while a position reference source J1056$+$7011 was used in the calibration for the other receiver. 

We used a task FRING to solve residual phase fluctuations of the reference source J1056$+$7011, and a task CALIB was also used to solve shorter phase fluctuation. 
Then, the solutions of phase, rate and amplitude were applied to the data of R~UMa. 
The instrumental phase difference between the two receivers, which was measured during each observation using the correlated data of noise signal injected into the two receivers from artificial noise sources installed on a feedome base of the VERA antenna \citep{hon08}, was also applied in the reduction process. 
Since the $\it{a~priori}$ delay tracking model used in the Mitaka correlator was not accurate enough for precise astrometry, we calibrated them based on more accurate delay tracking model. 
We changed it to the newly calculated model with CALC3/MSOLV \citep{jik05,man91} software package. 
The new recalculated model contains an estimation of the wet component of atmosphere from GPS data at each VERA station \citep{hon08}.
We made the images of maser spots associated with R~UMa using a task IMAGR. 
The field of view of each image was 81.92 mas $\times$ 81.92 mas with a pixel size of 0.04 mas $\times$ 0.04 mas. 
With a typical integration time of 5.7 hours, rms noise levels of the phase-referenced images ranged from 0.06 Jy\,beam$^{-1}$ to 0.76 Jy\,beam$^{-1}$. 
As a detection criterion of the maser spot on phase-referenced images, we adopted a signal-to-noise ratio $(S/N)$ of 5. 
For each observation, we fitted an observed brightness distribution with a two-dimensional Gaussian model and adopted its peak as positions of the maser spot. 
Then, these positions are used to derive a parallax and linear proper motions of the detected maser spots. 

%●表　座標
\begin{table}[htbp]
\caption{Coordinates of the sources.}
\begin{center}
\label{coordinate}
{\small  
\begin{tabular}{lll} 
\hline 
Source      & RA (J2000.0)            & DEC (J2000.0)         \\ 
\hline \hline
R~UMa           & 10$^\mathrm{h}$ 44$^\mathrm{m}$ 38$^\mathrm{s}$.42831 & $+68^{\circ}$ 46$\arcmin$ 32$\arcsec$.3442 \\ 
J1056$+$7011& 10$^\mathrm{h}$ 56$^\mathrm{m}$ 53$^\mathrm{s}$.61517 & $+70^{\circ}$ 11$\arcmin$ 45$\arcsec$.91561 \\ \hline
\end{tabular}
} 
\end{center}
\end{table}

%●表_観測
\begin{table*}[htbp]
\caption{Status of the VLBI observations}
\label{obs_table}
\begin{center}
\begin{tabular}{crcccccc} 
\hline
Obs.& Date\,\,\,\,\,\,\,\, &MJD&Days from& \multicolumn{4}{c}{$T_{\mathrm{sys}}$ [K]} \\ 
\cline{2-2} \cline{5-8}
ID & Y\,\,\,\,\,\,\,\,M\,\,\,\,\,D\,&&2006/1/1& MIZ & IRK & OGA & ISG   \\ \hline\hline 
 1 & 2006 Apr 11 &53836&101&200\,--\,1000  & 200\,--\,1400 & 200\,--\,500 & 300\,--\,500 \\
 2 &         May 14 &53869&134&130\,--\,190   & 160\,--\,260  & 200\,--\,450 & 500\,--\,8700 \\
 3 &         Aug 13 &53960&225&300\,--\,13000 & 300\,--\,9800 & 400\,--\,3400 & 300\,--\,700 \\
 4 &         Aug 30 &53977&242&200\,--\,400   & 400\,--\,13000 & 300\,--\,3000 & 400\,--\,9000\\
 5 &         Sep 12 &53988&255&300\,--\,600   & 400\,--\,13000 & 300\,--\,700 & 400\,--\,2000\\
 6 &         Oct 13 &54021&286&150\,--\,200   & 150\,--\,200 & 500\,--\,15000 & 300\,--\,800\\
 7 &         Oct 29 &54037&302&150\,--\,200   & 150\,--\,200 & 400\,--\,800 & 400\,--\,1200\\
 8 &         Nov 23 &54062&327&100\,--\,150   & 200\,--\,1000 & 300\,--\,700 & 300\,--\,1800\\
 9 &         Dec 21 &54090&355&100\,--\,150   & 100\,--\,150 & 400\,--\,900 & 200\,--\,2300\\ \hline 
10 & 2007 Jan 20 &54120&385&100\,--\,150   & 200\,--\,500 & 200\,--\,450 & 300\,--\,12000\\
11 &        Feb 23 &54154&419&100\,--\,150   & 100\,--\,150 & 150\,--\,12000 & 200\,--\,400\\
12 &         Apr 04 &54194&459&100\,--\,150   & 100\,--\,200 & 150\,--\,300 & 200\,--\,500\\
13 &         May 09 &54229&494&150\,--\,250   & 100\,--\,200 & 300\,--\,1800 & 200\,--\,350\\
14 &         Sep 11 &54354&619&200\,--\,1400  & 200\,--\,400 & 300\,--\,450 & 400\,--\,800\\
15 &         Oct 14 &54387&652&150\,--\,200   & 150\,--\,250 & $\cdots$ & 350\,--\,800\\
16 &         Nov 13 &54417&682&150\,--\,200   & 200\,--\,300 & 150\,--\,300 & 200\,--\,1000\\
17 &         Dec 14 &54448&713&100\,--\,200   & 150\,--\,300 & 150\,--\,300 & 200\,--\,800\\ \hline
18 & 2008  Jan 11 &54476&741&200\,--\,700   & 200\,--\,750 & 200\,--\,400 & 300\,--\,800\\
19 &         Feb 11 &54507&772&100\,--\,200   & 200\,--\,500 & 200\,--\,300 & 300\,--\,2200\\
20 &         Mar 08 &54533&798&100\,--\,150   & 100\,--\,200 & 150\,--\,200 & 200\,--\,500\\
21 &         Apr 02 &54558&823&100\,--\,200   & 200\,--\,900 & 150\,--\,200 & 500\,--\,4000\\
22 &         May 05 &54591&856&300\,--\,4000  & 150\,--\,400 & 200\,--\,300 & 400\,--\,8000\\
23 &         Jun 01 &54618&883&200\,--\,400   & 200\,--\,600 & 400\,--\,12000 & 400\,--\,5000\\
24 &         Jul 06  &54653&918&300\,--\,600   & 300\,--\,800 & 300\,--\,450 & 400\,--\,2000\\
\hline
\end{tabular} 
\end{center}  
\end{table*}  

%●Observationとしての単一鏡観測の詳細
\subsection{Single-dish observations}
In parallel with the VLBI observations, a total power spectrum of H$_2$O maser has been monitored at the VERA Iriki station with a typical interval of one month in a single-dish monitoring program that started in 2003 September  \citep{shi08}. 
The conversion factor from the antenna temperature to the flux density is about 20\,Jy\,K$^{-1}$.
The 1-$\sigma$ noise level of the single-dish observations is 0.05 K, corresponding to $\sim$1\,Jy, in integration time of 10 to 30 min. 

%●赤外線の観測と解析
\subsection{Near-infrared observations}
We also carried out near-infrared observations of R~UMa from January 2007 to May 2009 using Kagoshima University 1m telescope. 
The near-infrared camera equipped with the telescope has a 512$\times$512 pixels HAWAII arrays which provides the 
$J$-, $H$-, and $K$-band images. 
An image scale of the array is 0$'$$'$.636 pixel$^{-1}$, yielding a field of view of 5$'$.025$\times$5$'$.025. 

The data reduction and photometry for these data were carried out using the National Optical Astronomy Observatory's Imaging Reduction and Facility (IRAF) software package. 
Standard procedures of data reduction were adopted. 
Subtracting the average dark frame, normalizing each of the dark-subtracted images by the flat-field frame, then the sky frame was subtracted from the normalized image. 
The photometry was carried out with the IRAF/APPHOT package. 
Since R~UMa is bright in near-infrared and it causes a saturation of the detector, we used defocused images for the photometry. 

%●%%%%%%%%%%%%%%%%%%%%%%% Results %%%%%%%%%%%%%%%%%%%%%%%%%%%%%%%
\section{Results}
%%%%%%%%%%%%%%%%%%%%%%%%%%%%%%%%%%%%%%%%%%%%%%%%%%%%%%%%%%%
%●Results入来での単一鏡観測結果
\subsection{Single-dish results at Iriki station}
\label{singledish}
\begin{figure}[htbp]
\begin{center}
\includegraphics[width=85mm, angle=0]{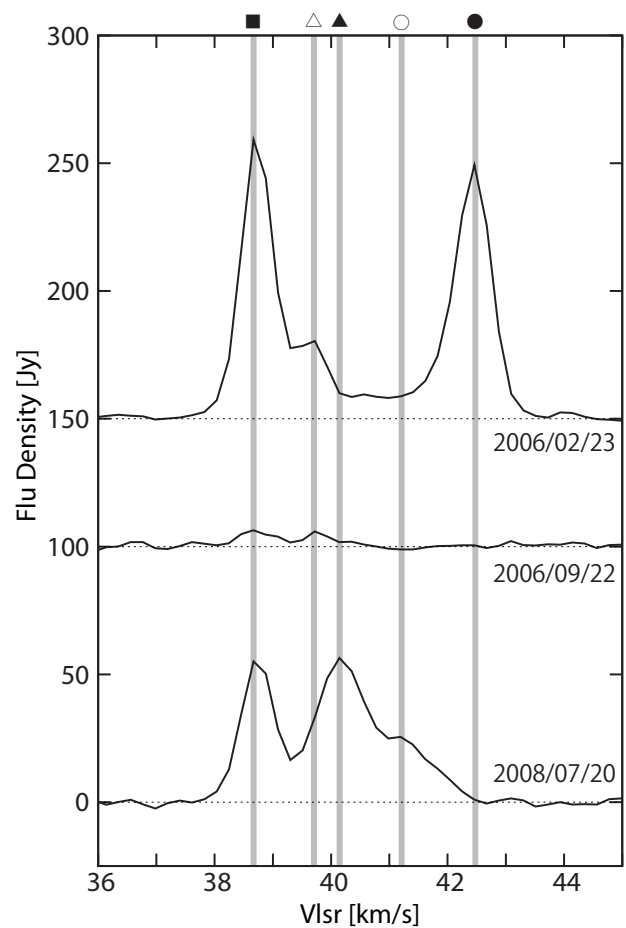} 
\end{center}
\caption{
Total power spectra of H$_2$O maser in R~UMa observed at VERA Iriki station in 2006 Feb 23, Sep 22, and 2008 Jul 20. 
Vertical grey lines and symbols indicate five major velocity components at $V_{\mathrm{LSR}}$ of 38.7 (filled square), 39.7 (open triangle), 40.1 (filled triangle), 41.1 (open circle), and 42.4 (filled circle)\,km\,s$^{-1}$, respectively. 
Noise floors of each spectrum are presented with dotted horizontal lines.
}
\label{fig-spectr}
\end{figure}

\begin{figure}[htbp]
\begin{center}
\includegraphics[width=85mm, angle=0]{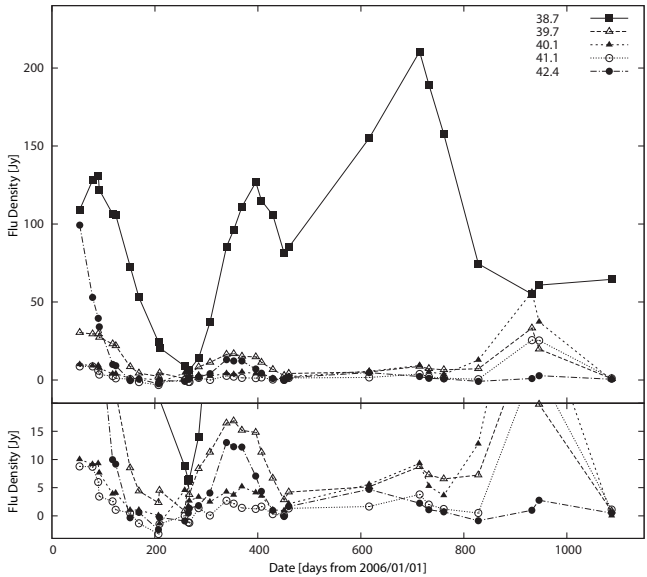} 
\end{center}
\caption{
Time variation of H$_2$O maser emissions of the five major velocity components from 2006 Jan 1.  
Same symbols as in figure~\ref{fig-spectr} are used to denote each velocity. 
Variation of weak components of 0 to 20\,Jy is magnified in the lower panel.  
}
\label{maser_time_var}
\end{figure}

From the single-dish monitoring at Iriki, we obtained spectral profile and time variation of H$_2$O maser emissions of R~UMa. 
Figure~\ref{fig-spectr} shows a total power spectra obtained on 2006 Feb 23 (top),  Sep 22 (middle) and 2008 Jul 20 (bottom), showing five major velocity components. 
Noise floors of each spectrum are shifted by 150 and 100 Jy. 
Figure~\ref{maser_time_var} shows a time variation of the five components at LSR velocity ($V_{\mathrm{LSR}}$) of 38.7 (filled square), 39.7 (open triangle), 40.1 (filled triangle), 41.1 (open circle), and 42.4 (filled circle)\,km\,s$^{-1}$ from 2006 to 2008.  
A velocity component at $V_{\mathrm{LSR}}$ of 38.7 km\,s$^{-1}$ has been dominant among all velocity components during our monitoring period. 
Same symbols are used in figure~\ref{fig-spectr} and figure~\ref{maser_time_var} to denote each velocity component. 

%●図　スペクトル

%●図　メーザー強度の時間変化

%%%%%%%%%%%%%%%%%%%%%%%%%%%%%%%%%%%%%%%%%%%%%%%%%%%%%%%%%%%
%●サブセクション：Results年周視差と距離
\subsection{Astrometric error estimation} 
\label{sec_error}
Position errors of a maser spots consist of several factors in phase-referencing VLBI \citep{hon07, nak08}. 
Before we derive a parallax of the source, we mainly consider two error factors in this section, an atmospheric effect and signal-to-noise ratio of the map. 
At first, we consider a contribution from wet zenith excess path. 
From data reduction of our previous study \citep{nak08}, we derived a typical zenith excess path length to be 30 mm. 
Using the same procedure in section~4 of \citet{hon07}, we estimated the position error of 0.08 mas due to this atmospheric effect. 
Separation angle (1.78$^\circ$) of the source pair causes a path-length error of 0.93 mm ($=$\,30 mm$\times$1.78$^{\circ}$/57$^{\circ}$.3/rad), which causes a position error of 0.08 mas ($=$ 0.93 mm/2.3$\times 10^9$ mm where 2.3$\times 10^9$ mm is the maximum baseline length of the VERA). 
Next, we consider an error which depends on a signal-to-noise ratio ($S/N$) of the phase-referenced map. 
Throughout our series of VLBI observations, major and minor axes of a typical synthesized beam are 1.2 mas $\times$ 0.8 mas with a position angle of 155$^\circ$. 
Projection of the synthesized beam onto RA and DEC axes gives $\theta_{\rm{b}}^{\rm{x}}=$0.73 mas and $\theta_{\rm{b}}^{\rm{y}}=$ 1.09 mas, respectively. 
We estimated this error contributions as the beam size devided by $S/N$ values. 
Since the $S/N$ varies from one observation to the next and from one maser spot to the next, we applied their respective $S/N$ values in the parallax fitting process. 
On the phase-referenced map with high $S/N$ value, the corresponding error was estimated to be, for example, 0.09\,mas in RA axis. 
Conversely, the error was estimated to be 0.22\,mas in DEC axis on the map indicating low $S/N$ value. 
Although the antenna position errors can be another error source in the astrometry, this contribution is one order of magnitude smaller than the others. 
These position errors were considered in the parallax fitting which is given in the next section.  

%%%%%%%%%%%%%%%%%%%%%%%%%%%%%%%%%%%%%%%%%%%%%%%%%%%%%%%%%%%%%%%%%%%%%%%%%%%%%%%%%%%%%%%%%%%%%%%%%%%%%%%%%%%%%%%%%%%%%%%%%%%%%%%%%%%%%%%%%%%%%%%%%%%%%%%%%%%%%%%%%%%%%%%%%%%%%%%%%
%●サブセクション：Results年周視差と距離
\subsection{Annual parallax and distance of R~UMa} 
\label{sec_parallax}
From the phase-referencing analysis, we detected H$_2$O maser spots in 18 observation among the total of 24 observations. 
In the data reduction of six observations on 2006 Aug 13, Aug 30, Sep 12, 2007 Sep 11, 2008 Jan 11, and Jul 6, we failed to detect maser spots and we could not use these data sets to estimate the parallax. 
On 2006 Aug 13 and 30, the maser emission was so faint that we could not detect them. 
Phase-referencing mapping of the data of 2006 Sep 12 was unable because of lacks of wet atmosphere values which is necessary for data reduction. 
The data recorded on 2007 Sep 11 and 2008 Jan 11 were invalid because of unknown troubles in data acquisition process, therefore we could not obtain any valid solutions from the data. 
There is a failure in frequency setting of the observation on 2008 Jul 6, and we failed to make phase-referencing map of this data. 
Detected maser spots represent $V_{\mathrm{LSR}}$ range between 38.17 and 42.81km\,s$^{-1}$, corresponds to 23 discrete velocity channels in our data acquisition setting.  
In table~\ref{detect_table}, we summarized detection flags of maser spots for each observation using same observation IDs in table~\ref{obs_table}. 
The flags ``0" and ``1" indicate non-detection and detection at corresponding velocity channel. 

Although the maser spots in many velocity channels show stable point-like structures on phase-referenced maps, structures at $V_{\mathrm{LSR}}$ of 38.38, 38.59, 38.80, and 39.44 km\,s$^{-1}$ are complicated and unstable. 
This brings a difficulty for an identification of the maser spots and we excluded them from parallax estimation.
Maser spots at seven more velocity channels were also excluded because the observation number of successful detection (flag ``1") is very limited and it is difficult to justify the parallax fitting. 
In table~\ref{detect_table}, $V_{\mathrm{LSR}}$ of channels not considered in the parallax fitting were presented with daggers ($\dag$). 

\begin{figure}
\begin{center}
 \includegraphics[width=85mm, angle=0]{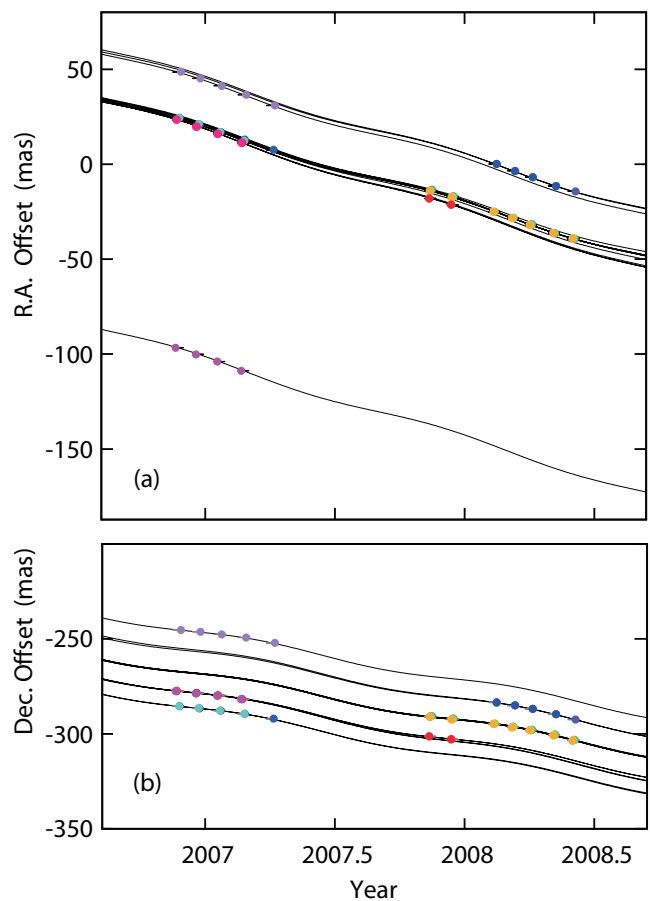} 
\end{center}
\caption{
Position offset of the maser spot in R~UMa in RA\,(a) and DEC.\,(b) along with date. 
Filled circles and solid lines represent observation data and best fit models, respectively. 
}
\label{fig_prop}
\end{figure}

\begin{figure}
\begin{center}
 \includegraphics[width=85mm, angle=0]{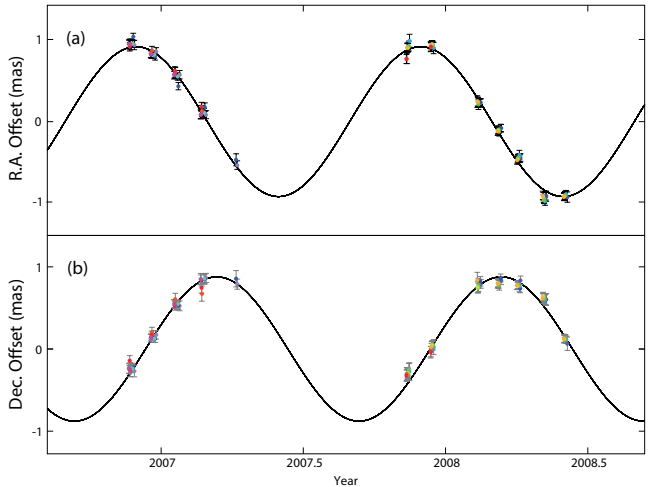} 
\end{center}
\caption{
Parallactic oscillation along the axes of RA (a) and DEC (b) of the maser spots. 
Filled circles and solid lines represent observation data and best fit models, respectively. 
}
\label{fig_parallax}
\end{figure}

Even in a single velocity channel, there are sometimes several spatially discrete maser spots that show stable structures. 
As a result, we identified 15 velocity channels and detected the total of 82 maser spots throughout the series of observations. 
They provide 164 numerical values (RA and DEC positions of each spot). 
Using these observable values, we solved for 61 free parameters which is consist of individual proper motions and initial positions of 15 maser spots along RA and DEC axes, and one common annual parallax. 
From a least-squares analysis, a parallax was obtained to be 1.97$\pm$0.05\,mas, corresponds to the distance of 508$\pm$13\,pc.  
A reduced-chi square $( = \chi^2/\nu)$ of this fitting was calculated to be 1.97, where $\chi^2$ is the squared sum of the residuals of fitting over the square of the uncertainty of each data point, and $\nu ( = 164 - 61 = 103 )$ is the degree of freedom. 
The obtained proper motions will be used for an analysis of the internal motions given in section~\ref{sec_internalmotion}. 
Figure~\ref{fig_prop} (a) and (b) represent motions of the 15 maser spots in RA and DEC axes with respect to defined coordinate of the star. 
We can see westward and southward systemic motions for all maser spots modulated by parallactic oscillation.  
In figure~\ref{fig_parallax}, we present oscillating terms of the parallactic motion in RA (a) and DEC (b).
In both figures, filled circles represent our observation data with error bars, and color denotes velocity, solid lines indicate the best fit models. 
Symbols in same observation date are plotted with slight shifts along time axis to avoid their overlap.

%●18列の位相補償星取表
\begin{table*}
\begin{center}
\caption{Maser detection on phase referenced maps.}
\label{detect_table}
\begin{tabular}{lp{1.5mm}p{1.5mm}p{1.5mm}p{1.5mm}p{1.5mm}p{1.5mm}p{1.5mm}p{1.5mm}p{1.5mm}p{1.5mm}p{1.5mm}p{1.5mm}p{1.5mm}p{1.5mm}p{1.5mm}p{1.5mm}p{1.5mm}p{1.5mm}}
\hline 
 & \multicolumn{18}{c}{Obs. ID} \\ \cline{2-19}　$V_{\mathrm{LSR}}$ & & & & & & & & & & & & & & & & & \\ 

[km\,s$^{-1}$] & 1 & 2 & 6 & 7 & 8 & 9 & 10 & 11 & 12 & 13 & 15 & 16 & 17 & 19 & 20 & 21 & 22 & 23 \\
\hline \hline 
42.81\footnotemark[$\dag$] & 1 & 0 & 0 & 0 & 1 & 1 & 1 & 0 & 0 & 0 & 0 & 0 & 0 & 0 & 0 & 0 & 0 & 0 \\
42.60                              & 1 & 0 & 0 & 0 & 1 & 1 & 1 & 1 & 0 & 0 & 0 & 0 & 0 & 0 & 0 & 0 & 0 & 0 \\
42.39                              & 1 & 1 & 0 & 0 & 1 & 1 & 1 & 1 & 0 & 0 & 1 & 0 & 0 & 0 & 0 & 0 & 0 & 0 \\
42.18                              & 1 & 0 & 0 & 0 & 1 & 1 & 1 & 1 & 0 & 0 & 1 & 1 & 1 & 0 & 0 & 0 & 0 & 0 \\
41.97                              & 0 & 0 & 0 & 0 & 1 & 1 & 1 & 1 & 0 & 0 & 1 & 1 & 1 & 0 & 0 & 0 & 0 & 0 \\
41.76\footnotemark[$\dag$] & 0 & 0 & 0 & 0 & 0 & 0 & 0 & 0 & 0 & 0 & 1 & 1 & 1 & 0 & 0 & 0 & 0 & 0 \\
41.54\footnotemark[$\dag$] & 0 & 0 & 0 & 0 & 0 & 0 & 0 & 0 & 0 & 0 & 1 & 1 & 1 & 0 & 0 & 0 & 0 & 1 \\
41.33\footnotemark[$\dag$] & 0 & 0 & 0 & 0 & 0 & 0 & 0 & 0 & 0 & 0 & 0 & 0 & 0 & 0 & 0 & 0 & 0 & 1 \\
41.12\footnotemark[$\dag$] & 0 & 0 & 0 & 0 & 0 & 0 & 0 & 0 & 0 & 0 & 1 & 0 & 0 & 0 & 0 & 0 & 0 & 1 \\
40.91\footnotemark[$\dag$] & 0 & 0 & 0 & 0 & 0 & 0 & 0 & 0 & 0 & 0 & 1 & 1 & 1 & 0 & 0 & 0 & 0 & 1 \\
40.70                              & 0 & 0 & 0 & 0 & 0 & 0 & 0 & 0 & 0 & 0 & 1 & 1 & 1 & 1 & 1 & 1 & 1 & 1 \\
40.49                              & 0 & 0 & 0 & 0 & 0 & 0 & 0 & 0 & 0 & 0 & 1 & 1 & 1 & 1 & 1 & 1 & 1 & 1 \\
40.28                              & 0 & 0 & 0 & 0 & 0 & 0 & 0 & 0 & 0 & 0 & 1 & 1 & 1 & 1 & 1 & 1 & 1 & 1 \\
40.07                              & 1 & 1 & 0 & 1 & 1 & 1 & 1 & 0 & 0 & 0 & 1 & 1 & 1 & 1 & 1 & 1 & 1 & 1 \\
39.86                              & 1 & 1 & 1 & 1 & 1 & 1 & 1 & 1 & 0 & 0 & 1 & 1 & 1 & 0 & 1 & 1 & 1 & 1 \\
39.65                              & 1 & 1 & 1 & 1 & 1 & 1 & 1 & 1 & 1 & 0 & 1 & 1 & 1 & 0 & 0 & 1 & 1 & 1 \\
39.44\footnotemark[$\dag$] & 1 & 1 & 1 & 1 & 1 & 1 & 1 & 1 & 1 & 0 & 0 & 0 & 1 & 1 & 1 & 0 & 0 & 1 \\
39.23                              & 1 & 1 & 0 & 0 & 0 & 0 & 0 & 0 & 0 & 0 & 0 & 0 & 1 & 1 & 1 & 1 & 1 & 1 \\
39.02                              & 0 & 0 & 0 & 0 & 1 & 1 & 1 & 1 & 1 & 0 & 1 & 1 & 1 & 1 & 1 & 1 & 1 & 1 \\
38.80\footnotemark[$\dag$] & 1 & 1 & 0 & 1 & 1 & 1 & 1 & 1 & 1 & 1 & 1 & 1 & 1 & 1 & 1 & 1 & 1 & 1 \\
38.59\footnotemark[$\dag$] & 1 & 1 & 0 & 1 & 1 & 1 & 1 & 1 & 1 & 1 & 1 & 1 & 1 & 1 & 1 & 1 & 1 & 1 \\
38.38\footnotemark[$\dag$] & 0 & 0 & 0 & 1 & 1 & 1 & 1 & 1 & 1 & 0 & 1 & 1 & 1 & 1 & 1 & 1 & 0 & 0 \\
38.17\footnotemark[$\dag$] & 0 & 0 & 0 & 0 & 0 & 0 & 0 & 1 & 0 & 0 & 1 & 0 & 0 & 0 & 0 & 0 & 0 & 0 \\
\hline
  \multicolumn{18}{@{}l@{}}{\hbox to 0pt{\parbox{110mm}{\footnotesize
  \smallskip
 \par\noindent
 \footnotemark[$\dag$] Velocity channels not used in parallax estimation.
     }\hss}}
\end{tabular} 
\end{center}  
\end{table*}

%●図：固有運動＋年周視差

%●図：年周視差振動成分

%%%%%%%%%%%%%%%%%%%%%%%%%%%%%%%%%%%%%%%%%%%%%%%%%%%%%%%%%%%
%●サブセクション：星周メーザーの分布 
\subsection{Distribution of circumstellar masers} 
In table~\ref{maser_table}, we presented all maser spots detected in our phase-referencing analysis. 
In the $V_{\mathrm{LSR}}$ range from 38.17 to 42.81 km\,s$^{-1}$, we identified 72 maser spots. 
Each maser spots moves in accordance with a systemic motion of R~UMa. 
In order to reveal a circumstellar distribution of maser spots on 2006 April 11 (date of the first observation), we calculated their positions by extrapolating initially detected positions using the R~UMa's systemic motion in the revised Hipparcos catalog ; {\boldmath $\mu$}$^{\mathrm{HIP}}$ = ($-40.51\pm0.79, -22.66\pm0.78$) mas\,yr$^{-1}$ \citep{van07}. 
Though the Hipparcos measurements have errors, we assumed no error in this calculation. 
Obtained maser spot positions in RA and DEC ($X, Y$) are presented in third and fourth columns in table~\ref{maser_table}. 
Since flux densities ($S$) and signal-to-noise ratios ($S/N$) vary with time, we used these values at their first detection. 

\begin{figure}
\begin{center}
\includegraphics[width=85mm, angle=0]{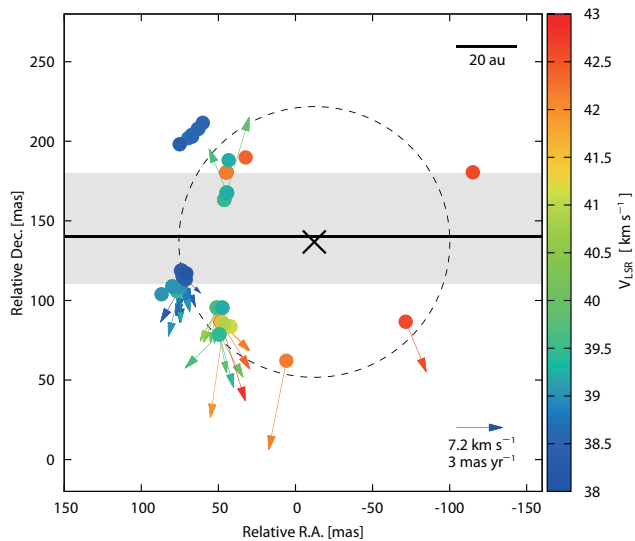}
\end{center}
\caption{
Distribution and internal motion of maser spots in R~UMa. 
Map angular size of 300\,mas square corresponds to 152 au square at the source distance of 508 pc. 
Filled circles indicate maser spots, and its color indicates $V_{\mathrm{LSR}}$ from 38 to 43\,km\,s$^{-1}$.  
Horizontal line with peripheral region and cross represent possible positions of the central star. 
Shell radius of 85 mas is presented with a dotted circle. 
See section~\ref{sec_stellarposition} for more details. 
}
\label{fig_internalmotion}
\end{figure}

Figure~\ref{fig_internalmotion} shows an obtained distribution of the maser spots in 300 mas square on April 11, 2006. 
This area corresponds to 152 au square at the source distance estimated by this work, i.e., 508\,pc. 
Filled circles are maser spots, and their colors indicate $V_{\mathrm{LSR}}$ between 38 km\,s$^{-1}$(blue) to 43 km\,s$^{-1}$(red). A color index is shown at right of the figure. 
The majority of maser spots are concentrated in the east of the map, and a few maser spots are detected in the west. 
Regarding the $V_{\mathrm{LSR}}$ property along the DEC axis, there is a large scale velocity gradient 
from blue- to red-shifted along the east to west direction. 
Details of arrows in figure~\ref{fig_internalmotion}, which indicate internal motions of the maser spots, will be given in section~\ref{sec_internalmotion}. 

%%● 図 メーザーの内部運動

%●表：スポットの内部運動
\begin{longtable}{cccccccccc} 
 \caption{Parameters of the detected masers.}
 \label{maser_table}
 \hline 
 ID &$V_{\mathrm{LSR}}$& $X$& $Y$& $S$& $S/N$ &$\mu^{\mathrm{int}}_{\mathrm{x}}$& $\sigma^{\mathrm{int}}_{\mu\mathrm{x}}$& $\mu^{\mathrm{int}}_{\mathrm{y}}$& $\sigma^{\mathrm{int}}_{\mu\mathrm{y}}$ \\ 
$i$ &[km\,s$^{-1}$]& [mas]& [mas]& [Jybeam$^{-1}$]& &[mas\,yr$^{-1}$]&  & [mas\,yr$^{-1}$]& \\ 
 \hline \hline
 \endhead
 \hline
 \endfoot
 \hline
 \multicolumn{10}{@{}l@{}}{\hbox to 0pt{\parbox{137mm}{\footnotesize
\smallskip
Column~(1)---Component ID.
Column~(2)---LSR velocity in km\,s$^{-1}$. 
Column~(3)---positions in RA relative to the original phase center. 
Column~(4)---positions in DEC relative to the original phase center.
Column~(5)---brightness of the spot at the first detection in Jy\,beam$^{-1}$. 
Column~(6)---signal-to-noise ratio ($S/N$). 
Column~(7)---best fit linear internal motion in RA in mas\,yr$^{-1}$.
Column~(8)---standard errors of the motion in RA 
Column~(9)---best fit linear internal motion in DEC in mas\,yr$^{-1}$.
Column~(10)---standard errors of the motion in DEC. 
 }}}
 \endlastfoot
%ID & Vlsr & X & Y & S & S/N & ux & erruxHIP & uy & erruyHIP \\
1 & 42.81 & 44.87 & 180.43 & 3.03 & 15.5 & $\cdots$ & $\cdots$ & $\cdots$ & $\cdots$ \\ 
2 & 42.81 & 48.47 & 86.77 & 0.79 & 7.3 & $-$1.59 & 1.30 & $-$4.99 & 0.79 \\ 
3 & 42.6 & $-$115.15 & 180.49 & 5.41 & 15.9 & $\cdots$ & $\cdots$ & $\cdots$ & $\cdots$ \\ 
4 & 42.6 & $-$71.55 & 86.61 & 1.67 & 13.2 & $-$1.34 & 0.86 & $-$3.16 & 0.81 \\ 
5 & 42.39 & 44.86 & 180.50 & 4.36 & 15.3 & $\cdots$ & $\cdots$ & $\cdots$ & $\cdots$ \\ 
6 & 42.39 & 32.27 & 189.81 & 0.43 & 3.9 & $\cdots$ & $\cdots$ & $\cdots$ & $\cdots$ \\ 
7 & 42.39 & 48.48 & 86.55 & 2.1 & 16.4 & $-$1.89 & 0.88 & $-$2.98 & 0.79 \\ 
8 & 42.18 & 44.82 & 180.57 & 1.99 & 11.8 & $\cdots$ & $\cdots$ & $\cdots$ & $\cdots$ \\ 
9 & 42.18 & 48.42 & 86.71 & 1.25 & 11.7 & $-$1.88 & 0.80 & $-$1.81 & 0.81 \\ 
10 & 42.18 & 5.86 & 61.99 & 1.36 & 10.4 & 1.15 & 0.79 & $-$5.58 & 0.78 \\ 
11 & 41.97 & 48.40 & 86.62 & 0.73 & 7.3 & $-$1.79 & 0.79 & $-$1.83 & 0.81 \\ 
12 & 41.97 & 46.42 & 84.72 & 1.5 & 7.9 & 0.87 & 0.79 & $-$5.86 & 0.78 \\ 
13 & 41.76 & 46.42 & 85.11 & 0.95 & 10.6 & $\cdots$ & $\cdots$ & $\cdots$ & $\cdots$ \\ 
14 & 41.54 & 47.67 & 86.17 & 0.7 & 4.9 & $\cdots$ & $\cdots$ & $\cdots$ & $\cdots$ \\ 
15 & 41.54 & 42.32 & 83.72 & 1.37 & 10 & $\cdots$ & $\cdots$ & $\cdots$ & $\cdots$ \\ 
16 & 41.33 & 42.28 & 83.62 & 1.93 & 15.7 & $\cdots$ & $\cdots$ & $\cdots$ & $\cdots$ \\ 
17 & 41.12 & 42.23 & 83.60 & 1.61 & 12.2 & $\cdots$ & $\cdots$ & $\cdots$ & $\cdots$ \\ 
18 & 40.91 & 50.30 & 94.94 & 1.83 & 7.4 & $\cdots$ & $\cdots$ & $\cdots$ & $\cdots$ \\ 
19 & 40.91 & 46.11 & 85.07 & 0.56 & 7.1 & $\cdots$ & $\cdots$ & $\cdots$ & $\cdots$ \\ 
20 & 40.91 & 47.50 & 94.99 & 0.95 & 7.3 & $\cdots$ & $\cdots$ & $\cdots$ & $\cdots$ \\ 
21 & 40.7 & 50.41 & 95.04 & 3.93 & 11.9 & 1.13 & 0.81 & $-$2.08 & 0.83 \\ 
22 & 40.49 & 50.42 & 95.09 & 4.88 & 12.3 & 1.14 & 0.81 & $-$2.05 & 0.84 \\ 
23 & 40.28 & 50.54 & 95.20 & 4.66 & 12.6 & 0.89 & 0.83 & $-$2.14 & 0.83 \\ 
24 & 40.07 & 44.62 & 167.91 & 1.41 & 9.5 & $\cdots$ & $\cdots$ & $\cdots$ & $\cdots$ \\ 
25 & 40.07 & 49.47 & 78.73 & 1.54 & 13.5 & $-$1.56 & 0.89 & $-$2.71 & 1.13 \\ 
26 & 40.07 & 50.74 & 95.28 & 3.37 & 12.5 & 0.45 & 0.84 & $-$2.20 & 0.83 \\ 
27 & 39.86 & 44.55 & 167.77 & 3.5 & 14.7 & $-$1.46 & 0.79 & 4.76 & 0.78 \\ 
28 & 39.86 & 46.24 & 163.17 & 1.5 & 7 & $\cdots$ & $\cdots$ & $\cdots$ & $\cdots$ \\ 
29 & 39.86 & 49.44 & 78.65 & 4.38 & 22.5 & $-$0.49 & 0.95 & $-$2.44 & 0.82 \\ 
30 & 39.86 & 50.89 & 95.45 & 1.86 & 8.6 & 0.16 & 0.83 & $-$2.25 & 0.86 \\ 
31 & 39.65 & 44.41 & 167.71 & 4.27 & 15.3 & 1.17 & 0.79 & 2.76 & 0.78 \\ 
32 & 39.65 & 46.20 & 163.31 & 1.47 & 7.1 & $\cdots$ & $\cdots$ & $\cdots$ & $\cdots$ \\ 
33 & 39.65 & 49.39 & 78.63 & 4.76 & 24.1 & 2.23 & 1.22 & $-$2.13 & 0.81 \\ 
34 & 39.65 & 51.15 & 95.49 & 0.91 & 5.5 & $-$0.23 & 0.90 & $-$2.09 & 0.84 \\ 
35 & 39.44 & 44.35 & 167.57 & 2.34 & 11.1 & $\cdots$ & $\cdots$ & $\cdots$ & $\cdots$ \\ 
36 & 39.44 & 46.21 & 163.19 & 0.83 & 4.6 & $\cdots$ & $\cdots$ & $\cdots$ & $\cdots$ \\ 
37 & 39.44 & 49.40 & 78.61 & 1.79 & 13.9 & $-$0.94 & 1.43 & $-$3.35 & 1.10 \\ 
38 & 39.44 & 76.83 & 105.81 & 0.96 & 8.8 & $\cdots$ & $\cdots$ & $\cdots$ & $\cdots$ \\ 
39 & 39.44 & 47.48 & 95.53 & 3.19 & 23.2 & $\cdots$ & $\cdots$ & $\cdots$ & $\cdots$ \\ 
40 & 39.23 & 44.32 & 167.28 & 1 & 7.6 & $\cdots$ & $\cdots$ & $\cdots$ & $\cdots$ \\ 
41 & 39.23 & 43.21 & 188.00 & 0.46 & 4.4 & $\cdots$ & $\cdots$ & $\cdots$ & $\cdots$ \\ 
42 & 39.23 & 78.81 & 108.18 & 2.06 & 14.4 & $\cdots$ & $\cdots$ & $\cdots$ & $\cdots$ \\ 
43 & 39.23 & 74.58 & 108.22 & 1.93 & 13.7 & $-$0.02 & 1.14 & $-$2.22 & 1.34 \\ 
44 & 39.23 & 47.45 & 95.47 & 0.99 & 8 & $\cdots$ & $\cdots$ & $\cdots$ & $\cdots$ \\ 
45 & 39.02 & 73.57 & 118.77 & 1.01 & 9.6 & $-$0.51 & 0.80 & $-$2.62 & 0.81 \\ 
46 & 39.02 & 86.94 & 103.90 & 1.91 & 5.2 & $\cdots$ & $\cdots$ & $\cdots$ & $\cdots$ \\ 
47 & 39.02 & 79.91 & 108.88 & 2.64 & 13.8 & $\cdots$ & $\cdots$ & $\cdots$ & $\cdots$ \\ 
48 & 39.02 & 74.54 & 108.32 & 3.03 & 16.7 & 0.79 & 0.95 & $-$3.10 & 0.97 \\ 
49 & 38.8 & 69.53 & 201.79 & 1.29 & 7.1 & $\cdots$ & $\cdots$ & $\cdots$ & $\cdots$ \\ 
50 & 38.8 & 63.10 & 207.75 & 0.99 & 8 & $\cdots$ & $\cdots$ & $\cdots$ & $\cdots$ \\ 
51 & 38.8 & 67.36 & 203.96 & 0.93 & 7.6 & $\cdots$ & $\cdots$ & $\cdots$ & $\cdots$ \\ 
52 & 38.8 & 73.86 & 118.84 & 2.18 & 12.7 & $-$0.46 & 0.95 & $-$1.88 & 1.04 \\ 
53 & 38.8 & 73.07 & 117.85 & 4.52 & 6.3 & 0.37 & 0.82 & $-$1.98 & 0.92 \\ 
54 & 38.8 & 73.16 & 117.80 & 1.63 & 9.5 & $-$0.18 & 0.85 & $-$2.03 & 0.84 \\ 
55 & 38.8 & 72.99 & 115.91 & 1.03 & 5.3 & $-$0.52 & 0.80 & $-$1.59 & 0.84 \\ 
56 & 38.59 & 60.03 & 211.63 & 1.27 & 6.3 & $\cdots$ & $\cdots$ & $\cdots$ & $\cdots$ \\ 
57 & 38.59 & 63.39 & 207.86 & 0.85 & 5.2 & $\cdots$ & $\cdots$ & $\cdots$ & $\cdots$ \\ 
58 & 38.59 & 67.08 & 202.93 & 1.13 & 8.3 & $\cdots$ & $\cdots$ & $\cdots$ & $\cdots$ \\ 
59 & 38.59 & 73.98 & 118.80 & 1.89 & 9.4 & $-$0.60 & 0.84 & $-$1.60 & 0.92 \\ 
60 & 38.59 & 72.50 & 116.98 & 7.39 & 12.9 & 0.28 & 0.82 & $-$2.23 & 0.97 \\ 
61 & 38.59 & 73.13 & 117.65 & 2.16 & 10.7 & $-$0.09 & 0.83 & $-$1.62 & 0.83 \\ 
62 & 38.59 & 72.23 & 115.53 & 4.72 & 6.2 & 0.48 & 0.83 & $-$2.52 & 0.92 \\ 
63 & 38.59 & 73.12 & 114.96 & 0.95 & 4.7 & $-$0.16 & 1.05 & $-$1.26 & 0.87 \\ 
64 & 38.59 & 72.01 & 114.43 & 4.13 & 10.1 & 0.56 & 0.97 & $-$2.42 & 0.88 \\ 
65 & 38.38 & 75.19 & 198.07 & 0.62 & 5.8 & $\cdots$ & $\cdots$ & $\cdots$ & $\cdots$ \\ 
66 & 38.38 & 74.19 & 118.79 & 0.73 & 6.1 & $-$0.93 & 0.92 & $-$2.24 & 0.88 \\ 
67 & 38.38 & 73.27 & 117.79 & 1.95 & 4.2 & 0.18 & 0.81 & $-$3.18 & 0.89 \\ 
68 & 38.38 & 73.20 & 117.60 & 0.92 & 7.6 & $-$0.33 & 0.91 & $-$2.27 & 0.81 \\ 
69 & 38.38 & 72.21 & 115.58 & 3.11 & 6.7 & 0.41 & 0.81 & $-$2.55 & 0.89 \\ 
70 & 38.38 & 72.95 & 114.83 & 0.57 & 4.8 & $-$1.14 & 1.04 & $-$1.01 & 0.97 \\ 
71 & 38.38 & 71.24 & 113.29 & 2.51 & 5.4 & 1.64 & 0.90 & $-$2.70 & 0.84 \\ 
72 & 38.17 & 70.90 & 116.85 & 0.61 & 9.3 & $\cdots$ & $\cdots$ & $\cdots$ & $\cdots$ \\ 
\end{longtable}

%●光学観測の結果
\subsection{Infrared light curve}
We carried out near infrared monitoring observations using Kagoshima University 1m telescope. 
Results of our photometry are presented in table~\ref{table_mag_jhk}. 
Apparent magnitudes in $J$-, $H$-, and $K$-band at each observation are presented as $m_J$, $m_H$, and $m_K$ with their errors. 
Assuming a sinusoidal function of 
\begin{eqnarray}
m_{J/H/K} = \Delta m \sin (2\pi \frac{T}{P} + \theta) + m_0 \nonumber, 
\end{eqnarray}
we solved pulsation period $P$, amplitude $\Delta m$, and mean magnitude $m_0$. 
We also solved a voluntary phase $\theta$ to conduct the numerical fitting. 
Time is presented as $T$ in the formula. 

\begin{figure}[htpb]
\begin{center}
 \includegraphics[width=85mm, angle=0]{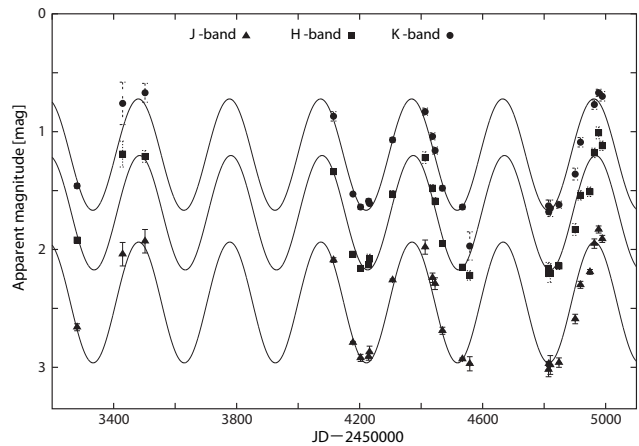}
\end{center}  
 \caption{
Infrared light curve of R~UMa in $J$- (triangle), $H$- (square), and $K$-bands (circle). 
Solid lines are best fit pulsation models in each band.
}
\label{lightcurve}
\end{figure}

In figure~\ref{lightcurve}, we show light curves in $J$- (triangle), $H$- (square), and $K$-band (circle) with best fit pulsation models. 
Three pulsation periods derived from the fitting are consistent within one day, and an averaged period was obtained to be 296.4 days. 
In $K$ band, mean magnitude and amplitude of pulsation were obtained to be $m_0=$ 1.19$\pm$0.02 mag and  $\Delta m =$ 0.47 mag, respectively. 
Other obtained parameters are summarized in table~\ref{lc_parm}. 

%●表：ＪＨＫバンド測光データ表
\begin{table*}[htbp]
\caption{$J$-, $H$-, $K$-band magnitudes.}
\label{table_mag_jhk}
\begin{center}
\begin{tabular}{cccc} 
\hline
MJD & $m_J$ & $m_H$ & $m_K$ \\ 
    & [mag] & [mag] & [mag] \\ \hline\hline 
53132 & 2.13$\pm$0.03 & 1.35$\pm$0.06 & 0.94$\pm$0.07 \\
53281 & 2.66$\pm$0.03 & 1.92$\pm$0.02 & 1.46$\pm$0.02 \\
53429 & 2.04$\pm$0.10 & 1.19$\pm$0.11 & 0.76$\pm$0.18 \\
53502 & 1.93$\pm$0.10 & 1.21$\pm$0.05 & 0.67$\pm$0.08 \\
54115 & 2.09$\pm$0.02 & 1.34$\pm$0.02 & 0.87$\pm$0.04 \\
54178 & 2.79$\pm$0.01 & 2.04$\pm$0.02 & 1.53$\pm$0.01 \\
54203 & 2.92$\pm$0.03 & 2.16$\pm$0.02 & 1.64$\pm$0.02 \\
54229 & 2.91$\pm$0.03 & 2.13$\pm$0.01 & 1.59$\pm$0.02 \\
54232 & 2.87$\pm$0.05 & 2.08$\pm$0.04 & 1.61$\pm$0.01 \\
54307 & 2.26$\pm$0.01 & 1.53$\pm$0.03 & 1.07$\pm$0.02 \\
54413 & 1.98$\pm$0.06 & 1.22$\pm$0.05 & 0.83$\pm$0.03 \\
54437 & 2.24$\pm$0.04 & 1.48$\pm$0.03 & 1.04$\pm$0.03 \\
54445 & 2.29$\pm$0.05 & 1.59$\pm$0.03 & 1.16$\pm$0.03 \\
54469 & 2.69$\pm$0.03 & 1.95$\pm$0.02 & 1.48$\pm$0.02 \\
54534 & 2.93$\pm$0.01 & 2.15$\pm$0.01 & 1.64$\pm$0.02 \\
54558 & 2.97$\pm$0.06 & 2.22$\pm$0.04 & 1.97$\pm$0.12 \\
54814 & 3.02$\pm$0.06 & 2.16$\pm$0.05 & 1.63$\pm$0.05 \\
54815 & 2.97$\pm$0.02 & 2.20$\pm$0.03 & 1.68$\pm$0.02 \\
54820 & 2.98$\pm$0.08 & 2.20$\pm$0.08 & 1.65$\pm$0.07 \\
54848 & 2.96$\pm$0.04 & 2.14$\pm$0.03 & 1.62$\pm$0.03 \\
54901 & 2.59$\pm$0.04 & 1.83$\pm$0.05 & 1.36$\pm$0.05 \\
54918 & 2.30$\pm$0.03 & 1.54$\pm$0.04 & 1.09$\pm$0.04 \\
54949 & 2.19$\pm$0.02 & 1.51$\pm$0.04 & --- \\
54963 & 1.95$\pm$0.04 & 1.18$\pm$0.04 & 0.77$\pm$0.04 \\
54977 & 1.83$\pm$0.03 & 1.01$\pm$0.05 & 0.67$\pm$0.03 \\
54989 & 1.91$\pm$0.03 & 1.12$\pm$0.04 & 0.70$\pm$0.04 \\
\hline
\end{tabular} 
\end{center}  
\end{table*}   

%●図：ＪＨＫバンドライトカーブ

%●表：ＪＨＫバンドライトカーブパラメーター表
\begin{table}[htbp]
\caption{Infrared light curve parameters.}
\label{lc_parm}
\begin{center}
\begin{tabular}{cccc} 
\hline
Band & $P$ & $\Delta m$ & $m_0$  \\
     &  [day] & [mag]     & [mag]    \\ \hline \hline
$J$ & 296.3 & 0.51 & 2.45$\pm$0.02  \\
$H$ & 296.6 & 0.49 & 1.69$\pm$0.02  \\
$K$ & 296.2 & 0.47 & 1.19$\pm$0.02  \\ \hline 
Average & 296.4& & \\ 
\hline
  \multicolumn{4}{@{}l@{}}{\hbox to 0pt{\parbox{60mm}{\footnotesize
  \smallskip
 \par\noindent
$P$: Pulsation period, $\Delta m$: amplitude, $m_0$: mean magnitude. 
     }\hss}}
\end{tabular} 
\end{center}  
\end{table}   

%%%%%%%%%%%%%%%%%%%%%%  Discussion　%%%%%%%%%%%%%%%%%%%%%%%%%%%%
\section{Discussion} 
%%%%%%%%%%%%%%%%%%%%%%%%%%%%%%%%%%%%%%%%%%%%%%%%%%%%%%%%%%%%%%%%
%●サブセクション：メーザーの内部運動
\subsection{Internal motion of circumstellar masers}
\label{sec_internalmotion} 
If we can directly detect a central star with our VLBI method, it is easy to know internal motions of maser spots with respect to the central star. 
But it is difficult, and we have to introduce some reasonable assumption in order to estimate internal motions of the maser spots.  
For example, an isotropy of circumstellar kinematics was assumed in our previous study \citep{nak14}. 
In this section, we try to reveal the internal motions of the maser spots with a method using an astrometric measurement from Hipparcos satellite~\citep{per97}. 

Proper motions of each maser spot measured in our VLBI observations ({\boldmath $\mu$}$^{\mathrm{VERA}}$) involve various kinematics, such as the Galactic rotation, a systemic motion of the star, and their internal motions. 
A proper motion of R~UMa measured by Hipparcos ({\boldmath $\mu$}$^{\mathrm{HIP}}$) also include the same kinematics as {\boldmath $\mu$}$^{\mathrm{VERA}}$ except for internal motions of the maser spots. 
Therefore, a remainder of two measurements ({\boldmath $\mu$}$^{\mathrm{VERA}}$ $-$ {\boldmath $\mu$}$^{\mathrm{HIP}}$) should give internal motions of the maser spots on the rest frame fixed to the central star.  
In the revised Hipparcos catalog~\citep{van07}, the proper motion of R~UMa is reported to be 
{\boldmath $\mu$}$^{\mathrm{HIP}}$ 
$=(-40.51\pm0.79, -22.66\pm0.78)$ mas\,yr$^{-1}$. 
For maser spots detected more than two continuous epochs, we estimated their proper motions. 
Since the parallax of 1.97 mas determined in section~\ref{sec_parallax} is the same for all maser spots, we used the fixed parallax and re-fitted all the individual maser spots solving only for internal motion.
By subtracting the {\boldmath $\mu$}$^{\mathrm{HIP}}$ from the proper motions of each maser spot, we obtained internal motions $\mu^{\mathrm{int}}_{\mathrm{x}}$ and $\mu^{\mathrm{int}}_{\mathrm{y}}$ for 38 out of all 72 maser spots, and presented them in table~\ref{maser_table} in unit of mas\,yr$^{-1}$.
We also presented errors of the internal motions $\sigma^{\mathrm{int}}_{\mu\mathrm{x}}$ and $\sigma^{\mathrm{int}}_{\mu\mathrm{y}}$ on the table which are quadratic sums of two errors from Hipparcos catalog and our VLBI observation. 
Since measurement accuracy of proper motions in our VLBI observation is, on average, 0.2\,mas\,yr$^{-1}$, resultant errors are almost dominated by the error from Hipparcos measurement. 

A proper motion of R~UMa system can also be estimated from our VLBI observations. 
We averaged out all the proper motions of maser spots and obtain a motion of $-40.77\pm0.39$ mas\,yr$^{-1}$ and $-24.75\pm0.38$ mas\,yr$^{-1}$ in RA and DEC, respectively. 
In the RA, two proper motions from Hipparcos and VERA are consistent within their errors. 
In the DEC, however, there is a difference of $\sim$2 mas\,yr$^{-1}$ between two measurements. 
As shown in figure~\ref{fig_internalmotion}, we found unisotropic distribution of maser spots. 
We think the systemic motion derived from VERA possibly be biased due to this unisotropy. 
Therefore, in this section, we used the proper motion in Hipparcos catalog to inspect internal motions of the maser spots. 

In figure~\ref{fig_internalmotion}, internal motions of maser spots on sky plane are indicated with arrows. 
Proper motion of 1 mas\,yr$^{-1}$ corresponds to a transverse velocity of 2.41 km\,s$^{-1}$ at the source distance of 508 pc. 
An arrow at bottom right shows a proper motion of 3 mas\,yr$^{-1}$ ($=$ 7.22km\,s$^{-1}$). 
Average errors of internal motions are 0.90 mas\,yr$^{-1}$ and 0.88 mas\,yr$^{-1}$ in RA and DEC, respectively. 

Most $\mu^{\mathrm{int}}_{\mathrm{y}}$ show essentially larger velocities than their errors $\sigma^{\mathrm{int}}_{\mu\mathrm{y}}$, and therefore, we think our analysis gives an reliable picture of the internal motions especially along DEC axis. 
Maser spots at the southern area of the map show southward motions, and on the contrary, two spots at the northern area show northward motions. 
With regards to the RA axis, $\mu^{\mathrm{int}}_{\mathrm{x}}$ values are same as their errors, and this brings a large uncertainty to directions about the internal motions along RA axis. 
Nevertheless, we can conclude that there is no remarkable systemic motion along RA axis enough to be detected with this method.  
As a result, our analysis of internal motion revealed an outward motions with respect to the central region of the map. 
A root sum square of internal motions along two axes $( = \sqrt{ (\mu^{\mathrm{int}}_{\mathrm{x}})^2 + (\mu^{\mathrm{int}}_{\mathrm{y}})^2}\,)$ gives a transverse velocities on the skyplane. 
An average of the transverse velocities was obtained to be 6.61 km\,s$^{-1}$. 
In the next section, we will give a more detailed study to capture a global picture of the maser spots and the central star. 

Finally in this subsection, we mention a binarity of the R~UMa system. 
There are some AGB samples known as a binary stars, i.e., Mira AB system \citep{kar97}, R~Aqr \citep{wil81}, and so on.  
The R~UMa is one of possible samples of binary system.  
\citet{sah08} studied fluxes of some AGB stars at near-UV, far-UV, optical, and near-IR bands. 
They concluded that the far-UV excess likely results either directly from the presence of a hot binary companion or indirectly from a hot accretion disk around the companion. 
If the R~UMa forms a binary system and the binarity affects internal motions of maser spots, astrometric measurements of the maser spots include the binary motion. 
Then, a remainder {\boldmath $\mu$}$^{\mathrm{VERA}}$ $-$ {\boldmath $\mu$}$^{\mathrm{HIP}}$ still have some contribution from the binary motion. 
However, as presented above, we could not detect any systematic motion for all maser spots but detected expanding like motions. 
This can implies that there is no large binary effect that seriously disturb our consideration of internal motion. 
Therefore, we have not considered the binarity in this section. 

%%%%%%%%%%%%%%%%%%%%%%%%%%%%%%%%%%%%%%%%%%%%%%
%●サブセクション：星の位置
\subsection{Stellar position and 3D-picture of the maser spots}
\label{sec_stellarposition}
\begin{figure}
\begin{center}
\includegraphics[width=80mm, angle=0]{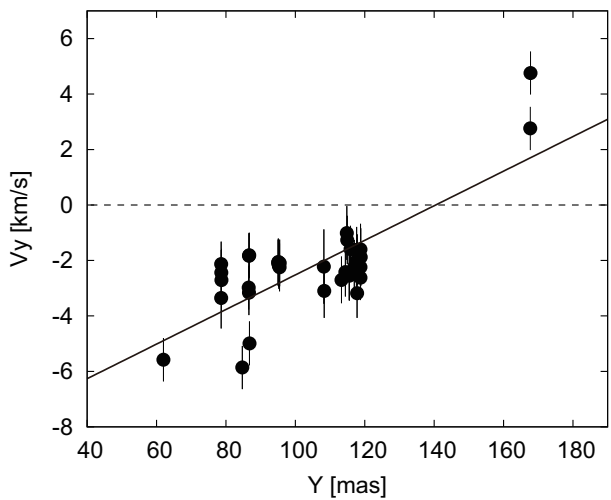}
\end{center}
\caption{
Velocity gradient of $V_{\mathrm{y}}$ along DEC ($Y$) axis. 
An intersection point of $V_{\mathrm{y}} = 0$ and the slope 
gives $Y = 140.45$ mas. 
}
\label{fig_yvsvy}
\end{figure}

Based on the distribution and internal motions of the maser spots revealed in previous sections, we will discuss the position of the central star. 
Now we focus on the motions of the maser spots along DEC axis. 
In figure~\ref{fig_yvsvy}, we show a relation between $\mu^{\mathrm{int}}_{\mathrm{y}}$ and $Y$ of 38 maser spots. 
Since there seems to be a gradient of $\mu^{\mathrm{int}}_{\mathrm{y}}$, we fitted this data to a linear function.
Then we obtained a relation of $\mu^{\mathrm{int}}_{\mathrm{y}} = 0.062 Y - 8.758$, which is presented with a solid line in figure~\ref{fig_yvsvy}. 
For $V_{\mathrm{y}} = 0$, this relation gives a $Y = 140.45$, and this $Y$ value can be considered as a possible $Y$ position of the central star. 
In figure~\ref{fig_internalmotion}, this $Y$ value is presented with a solid horizontal line with peripheral region (gray colored) indicating its error. 
Along the RA axis, it is difficult to find a gradient, we could not conduct the same analysis as the DEC axis. 

\begin{figure}
\begin{center}
\includegraphics[width=85mm, angle=0]{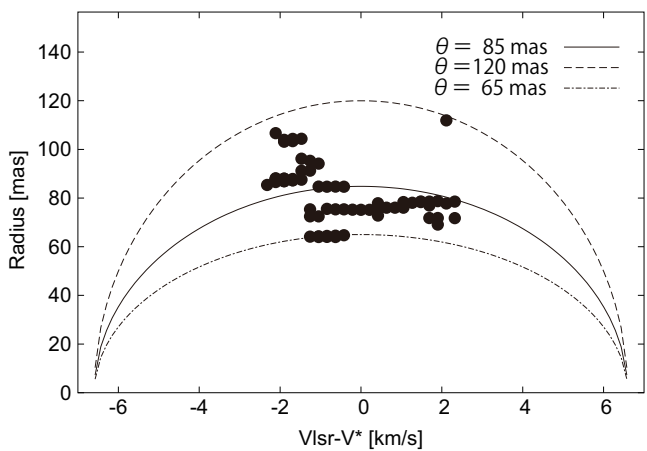}
\end{center}
\caption{
Relation between an angular radius $\theta$ and $V_\mathrm{LSR}-V_\ast$ of the maser spots. 
A solid ellipse indicates an uniform shell model with a radius of 85 mas obtained from a numerical fitting. 
Dashed and one-dotted-chain lines indicate models with shell radii of 120 and 65 mas, respectively.  
Expansion velocities of $V_\mathrm{exp} =$ 6.6 km\,s$^{-1}$ are fixed for all models. 
}
\label{fig_pv}
\end{figure}

Independently, we tried to estimate a stellar position using the angular distribution of maser spots. 
A cross on the map indicates an estimated stellar position of $(X, Y) = (-12.1, 136.8)$ mas which was obtained by simply calculating medians of two ends in RA and DEC axes. 
Assuming this position, we calculated angular distances $\theta$ of the maser spots from the central star. 
Here, we introduce a simple uniform expanding shell model \citep{oln77,rei77} of 
\begin{eqnarray}
\left(\frac{\theta}{\theta_\mathrm{m}}\right)^2 + \left(\frac{V_\mathrm{LSR}-V_\ast}{V_\mathrm{exp}}\right)^2 = 1,  \nonumber 
\end{eqnarray}
where $\theta$ is an angular distance from the central star, 
$\theta_\mathrm{m}$ is the shell radius, 
$V_\ast$ is the radial velocity of the star, and 
$V_\mathrm{exp}$ is an expansion velocity of the shell. 
Figure~\ref{fig_pv} shows a $\theta$ vs $(V_\mathrm{LSR}-V_\ast)$ diagram of the 38 maser spots. 
A shape of the model on this figure shows an ellipse. 
We assumed a $V_\ast$ of 40.49 km\,s$^{-1}$ which is obtained as a center of $V_\mathrm{LSR}$ values of maser spots. 
Since a distribution of the data on the $\theta - (V_\mathrm{LSR}-V_\ast)$ plane is too narrow to numerically solve suitable $V_\mathrm{exp}$ and $\theta_\mathrm{m}$, we assumed a $V_\mathrm{exp}$ to be 6.6 km\,s$^{-1}$ which is an average of transverse velocities of the maser spots obtained in previous section. 
Using all maser spots, we solved a $\theta_\mathrm{m}$ to be 85$\pm$2 mas, which is presented with solid line in figure~\ref{fig_pv}. 
For comparison, we also presented two models whose outer radius of 120 mas (dashed--line) and an inner radius of 65 mas (one--dotted chain line). 
Latter two models are not obtained from numerical fitting but we assumed fixed radii. 
In figure~\ref{fig_internalmotion}, the shell model with a radius of $\theta_\mathrm{m}=$ 85 mas is presented as a dotted circle whose center is assumed to be the cross symbol.  

%● 図 　Y vs Vy　の図

%● 図 　PV　の図

By revealing accurate motion of circumstellar matter, properties of stellar wind will be inspected and it helps us to understand mass-loss process. 
However, the distributions of circumstellar matter sometimes show symmetry, and sometimes show asymmetry. 
A conventional method for determining the circumstellar motion based on the VLBI data, which assumes uniformity or symmetry for circumstellar matters, intrinsically has limitation to derive  the real picture. 
Therefore, it is important to create different method which does not need any assumption about symmetry or uniformity for circumstellar matter. 
A capability of the new method, which combines VLBI and other independent astrometric data, was presented here.
To acquire more accurate picture of internal motions using the same method, accurate systemic motions of stars are required. 
The Gaia satellite~
\footnote[1]{Gaia Mission, ESA;  http://sci.esa.int/gaia/}, 
launched in 2013, is an ongoing astrometric program which expected to measure accurate proper motions for large amount of stars. 
For very bright stars, like nearby Mira variables, a Japanese satellite Nano--JASMINE~
\footnote[2]{Nano - JASMINE, NAOJ ;\\  http://www.jasmine-galaxy.org/index-en.html} 
will also be a powerful and promising telescope to determine their proper motions. 
In near future, more accurate proper motions of stars measured by new satellites will be tied up with VLBI measurements of maser spots, and circumstellar dynamics of many sources can be studied with the same method in this study. 
We expect a successful launch of Nano--JASMINE scheduled in near future. 

%%%%%%%%%%%%%%%%%%%%%%%%%%%%%%%%%%%%%%%%%%%%%%%%%%%%%%%%%%%
%●サブ： PLダイアグラム上の位置　%%%%%%%%
\subsection{R~UMa in the $M_K-$log$P$ Diagram}
\label{sec_plr}
For AGB stars in the LMC, it is known that there are several distinct sequences on the $m_K-$log$P$ diagram \citep{woo99, ita04-1}, where $m_K$ and $P$ indicate their $K$ band apparent magnitude and pulsation period. 
Sequences C and C' in their classification correspond to groups of variables pulsating in a fundamental tone and a first-overtone, respectively. 
Since distances of the Galactic sources can not be treated as identical like sources in the LMC, the same relation should be studied in $M_K-$log$P$ plane, where $M_K$ is a $K$ band absolute magnitude. 
In this section, we confirm a location of R~UMa on the $M_K-$log$P$ plane, and solve a zero-point of the Galactic $M_K-$log$P$ relation based on recent astrometric results. 

During the last decade, we have conducted astrometric observations for Mira and semiregular variables using VERA \citep{nak08}. 
Determination of $M_K = -8.33 \pm 0.10$ mag of U~Lyn by \citet{kam15} is the latest published result of Galactic Mira variable based on a parallax from our ongoing program. 
In this study, using our trigonometric distance of 508$\pm$13 pc and the infrared magnitude of $m_K = 1.19\pm0.02$ mag, we derived a $M_K = -7.34\pm0.06$\,mag for R~UMa. 
The error is determined as a root mean square of distance based error and apparent magnitude error. 
Adding this new source, we summarized the sources in table~\ref{parallax_table} in an increasing order of pulsation period together with all Galactic LPVs whose distances are determined with astrometric VLBI. 
%●SRについての記述
There are samples of Mira, semiregular (SRa, SRb in table~\ref{parallax_table}), and red supergiants (SRc in table~\ref{parallax_table}). 
Semiregular variables show amplitude smaller than Miras, and among them, stars with better periodicities are referred to as SRa to distinguish from SRb with poorly defined periodicities. 
Species of the masers used in the parallax measurements are also shown. 
References of the parallaxes and apparent magnitudes $m_K$ are given in the footnote to table~\ref{parallax_table}. 
Using the distance and $m_K$, we derived absolute magnitudes $M_K$ of the sources. 
In estimation of $M_K$ errors, only the distance errors were considered. 

\begin{figure}
\begin{center}
 \includegraphics[width=85mm, angle=0]{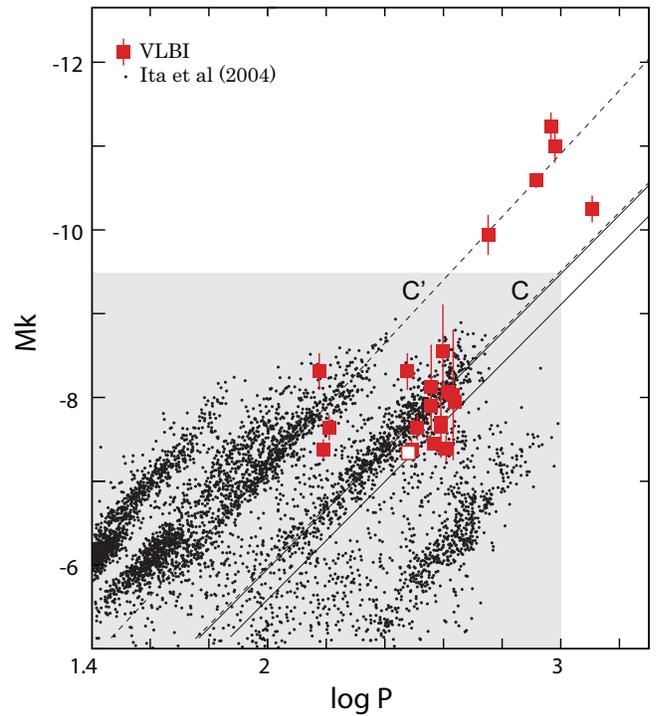}
\end{center}
\caption{
Absolute magnitudes ($M_K$) $-$ $\log P$ diagram of the Galactic long period variables. 
Filled red squares indicate sources in table~\ref{parallax_table}, whose distances are derived from astrometric VLBI. 
Only the result of R~UMa is presented with an open square. 
Solid lines show fitting results of $M_K - \log P$ relations for Galactic Mira variables in table~\ref{parallax_table}. 
See section~\ref{sec_plr} for detail of the fitting procedure. 
In a shaded area, LPVs in the LMC are presented with small dots \citep{ita04-1}.  
Labels C and C' and corresponding dashed 
lines indicate sequences in \citet{ita04-1}. 
}
\label{plr}
\end{figure}

Now, we define a $M_K-$log$P$ relation in the form of 
$M_K = -3.52 \, \mathrm{log}P + \delta$, where we assume a fixed slope of $-3.52$ determined by \citet{ita04-1} using LPVs in the LMC. 
Using Miras and four semiregular variables (RW~Lep, S~Crt, RX~Boo, and W~Hya) in table~\ref{parallax_table}, we solved the constant $\delta$. 
Unweighted and weighted least squares fitting to the data gives $\delta$ of 1.09$\pm$0.14 and 1.45$\pm$0.07, respectively. 
In \citet{kam12}, various periods of a well-studied semiregular source RX~Boo are reported, and ratios of the periods are found to be around two.  
In order to include all Mira and semiregular variables in the fitting, the periods of semiregular variables are multiplied by two and used. 
Red supergiants are not included in the fitting. 
Since $M_K$ errors of a few sources are quite small compared to other source, it gives a $\delta$ discrepancy of 0.36 between two fittings. 
In figure~\ref{plr}, we presented all sources in table~\ref{parallax_table} on $M_K-$log$P$ plane with red squares. 
To distinguish the current result of R~UMa from published ones, R~UMa is presented with an open square. 
Two solid lines indicate the results from unweighted (upper) and weighted (lower) fitting, respectively.
And, two dashed lines indicate relations derived by \citet{ita04-1} for sequence C' (first-overtone) and $C$ (fundamental tone), respectively. 
The LPVs in LMC reported by \citet{ita04-1} are presented with small dots in a shaded area of the figure. 
We can find that R~UMa falls on the sequence C, and it is likely that the star pulsates in a fundamental mode. 
To calibrate absolute magnitudes of the sources in LMC,  we assumed a distance modulus of 18.49 \citep{van07} for LMC. 
Within a determination accuracy of $\delta$ in our study, we find a consistency of the relations for Mira variables between our Galaxy and the LMC. 

In addition to Miras and semiregular variables, now we focus on four sources (3 red supergiants and NML~Cyg) in table~\ref{parallax_table}. 
They are $\sim$3 magnitudes brighter than Mira and semiregular variables, and show longer pulsation periods of 822 ($\log P = 2.915$) to 1280 ($\log P = 3.107$) days. 
If we extrapolate C and C' sequences of \citet{ita04-1} to brighter and longer period region, we find that three sources fall near C' and the other falls near C sequence, that possibly indicating a difference of the pulsation mode. 
Since the number of the Galactic red super giants with parallactic distance is small, it is important to increase the sample number for better calibration of their  $M_K-$log$P$ relation. 
The VLBI astrometry can be a useful method to provide their accurate distances. 

% 図PLR

%●年周視差の表（他の論文も含む、新形式）
\begin{table*}[!tb]
\caption{Results from VLBI astrometry}
\label{parallax_table}
\begin{center}
\begin{tabular}{lcccccccc} 
\hline
%%%%%% 2015 09 15  GCVS database 
Source & Type & Parallax & $P$ & $\mathrm{Log}P$ & $m_K$ & $M_K$ &Maser & Reference\footnotemark[$\dag$]  \\
       &      & [mas]          &[day]&           & [mag] & [mag] &      & (Parallax, $m_K$) \\ \hline \hline
RW~Lep & SRa   & 1.62$\pm$0.16   & 150 & 2.176 & 0.639 &    $-8.31\pm0.22$ &H$_2$O& kam14, a\\ 
S~Crt  & SRb  & 2.33$\pm$0.13    & 155 & 2.190 &    0.786&   $-7.38\pm0.12$ &H$_2$O& nak08, a\\ 
RX~Boo & SRb  & 7.31$\pm$0.5    & 162 & 2.210 & $-$1.96 &  $-7.64\pm0.15$ &H$_2$O& kam12, b\\ 
R~UMa  & Mira & 1.97$\pm$0.05  & 302 & 2.480 &  1.19  &  $-7.34\pm0.06$ &H$_2$O&  $\cdots$ \\ 

W~Hya  & SRa  &10.18$\pm$2.36   & 361 & 2.558 & $-$3.16 &  $-8.12\pm0.51$ &OH    & vle03, c\\
S~CrB  & Mira & 2.39$\pm$0.17   & 360 & 2.556 &    0.21   &   $-7.90\pm0.15$ &OH    & vle07, c\\
T~Lep  & Mira & 3.06$\pm$0.04    & 368 & 2.566 & 0.12&        $-7.45\pm0.03$ &H$_2$O& nak14, c\\ 
R~Aqr  & Mira & 4.7$\pm$0.8         & 390 & 2.591 & $-$1.01&   $-7.65\pm0.37$ &SiO   & kam10, c\\
R~Aqr  & Mira & 4.59$\pm$0.24      & 390 & 2.591 & $-$1.01&   $-7.70\pm0.11$ &SiO   & min14, c\\
RR~Aql & Mira & 1.58$\pm$0.40    & 396 & 2.598 &    0.46 &    $-8.55\pm0.56$ &OH    & vle07, c\\
U~Her  & Mira & 3.76$\pm$0.27    & 406 & 2.609 & $-$0.27 &   $-7.39\pm0.16$ &OH    & vle07, c\\
SY~Scl & Mira & 0.75$\pm$0.03     & 411 & 2.614 &    2.55&     $-8.07\pm0.09$ &H$_2$O& nyu11, b\\ 
R~Cas  & Mira & 5.67$\pm$1.95    & 430 & 2.633 & $-$1.80 &   $-8.03\pm0.78$ &OH    & vle03, c\\
U~Lyn  & Mira & 1.27$\pm$0.06     & 434 & 2.637 & 1.533 &     $-7.95\pm0.10$ &H$_2$O& kam15, a\\ 
UX~Cyg & Mira & 0.54$\pm$0.06   & 565 & 2.752 &    1.40   &   $-9.94\pm0.24$ &H$_2$O& kur05, a\\
S~Per & SRc & 0.413$\pm$0.017   & 822 & 2.915 & 1.33 &       $-10.59\pm0.09$ &H$_2$O& asa10, b\\
PZ~Cas & SRc &  0.356$\pm$0.026  & 925 & 2.966 & 1.00 &      $-11.24\pm0.16$ &H$_2$O& kus13, b\\
VY~CMa & SRc & 0.88$\pm$0.08   & 956 & 2.980 & $-$0.72 &  $-11.00\pm0.20$ &H$_2$O& cho08, b\\
NML~Cyg & --- & 0.62$\pm$0.047  & 1280 & 3.107 & 0.791 &   $-10.25\pm0.16$&H$_2$O & zha12, a\\
\hline
\hline 
\multicolumn{9}{@{}l@{}}{\hbox to 0pt{\parbox{145mm}{\footnotesize
\smallskip
\par\noindent
\footnotemark
[$\dag$]
References of the parallax are as follows : 
(kam14) \cite{kam14}, 
(nak08) \cite{nak08},
(kam12) \cite{kam12}, 
(vle03) \cite{vle03}, 
(vle07) \cite{vle07}, 
(nak14) \cite{nak14}, 
(kam10) \cite{kam10}, 
(min14) \cite{min14}, 
(nyu11) \cite{nyu11}, 
(kam15) \cite{kam15},   
(kur05) \cite{kur05}, 
(asa10) \cite{asa10},
(kus13) \cite{kus13}, 
(cho08) \cite{cho08}, and 
(zha12) \cite{zha12}. 
References of the apparent magnitudes ($m_K$) are as follows : 
(a) The IRSA 2MASS All-Sky Point Source Catalog \citep{cut03}, 
(b) Catalogue of Stellar Photometry in Johnson's 11-color system \citep{duc02}, 
(c) Photometry by \citet{whi00}. 
     }\hss}}
\end{tabular} 
\end{center}  
\end{table*} 

%%%%%%%%%%%%%%%%%%%%%%%%%%%%%%%%%%%%%%%%%%%%%%%%%%%%%%%%%%%
%● まとめ　%%%%%%%%
\subsection{Conclusion}
\label{sec_concl}
We conducted astrometric VLBI observation of a Mira variable R~UMa. 
Positions of H$_2$O maser spots at 22\,GHz were measured using the VERA array. 
Obtained parallax of 1.97$\pm$0.05 mas gives a distance of 508$\pm$13 pc. 
Circumstellar kinematics of 38 maser spots and angular distribution of all 72 maser spots were revealed from our observations, then we gave a constraint on a stellar position. 
The $M_K - \log P$ relations for the Galactic LPVs, whose distances were measured from astrometric VLBI, were explored and we obtained the relations of 
$M_K = -3.52 \, \mathrm{log}P + (1.09\pm0.14)$ and 
$M_K = -3.52 \, \mathrm{log}P + (1.45\pm0.07)$ 
from unweighted- and weighted-least squares fittings, respectively. 
R~UMa was found to fall on C sequence that pulsate in a fundamental tone. 
Positions of red supergiants in the same $M_K - \log P$ plane also found to fall on C' and C sequences. 
From this we can infer the pulsation mode of the sources, and also this represents an capability of VLBI astrometry for a calibration of $M_K - \log P$ relation applicable to red supergiants. 

%%%%%%%%%%%%%%%%%%%%%%%%%%%%%%%%%%%%%%%%%%%%%%%%%%%%%%%%%%%%%%%%%%%%%%%%%%%%%%%%%%%%%%%%%%%%%%%%%%%%%%%%%%%%%%%%%%%%%%%%%%%%%%%%%%%%%%%%%%%%%%%%%%%%%%%%%%%%%%%%%%%%%%%%%%
{}

\begin{thebibliography}{}
\bibitem[Asaki et al.(2010)]{asa10} Asaki, Y., Deguchi, S., Imai, H., et al.\ 2010, \apj, 721, 267 
\bibitem[Choi et al.(2008)]{cho08} Choi, Y.~K., Hirota, T., Honma, M., et al.\ 2008, \pasj, 60, 1007 
\bibitem[Cutri et al.(2003)]{cut03} Cutri, R.~M., Skrutskie, M.~F., van Dyk, S., et al.\ 2003, VizieR Online Data Catalog, 2246,  
\bibitem[Ducati(2002)]{duc02} Ducati, J.~R.\ 2002, VizieR Online Data Catalog, 2237
\bibitem[Feast et al.(1989)]{fea89} Feast, M.~W., Glass, I.~S., Whitelock, P.~A., \& Catchpole, R.~M.\ 1989, \mnras, 241, 375 
\bibitem[Gail \& Sedlmayr(2014)]{gai14} Gail, H.-P., \& Sedlmayr, E.\ 2014, Physics and Chemistry of Circumstellar Dust Shells, by Hans-Peter Gail , Erwin Sedlmayr, Cambridge, UK: Cambridge University Press, 2014
\bibitem[Habing \& Olofsson(2003)]{hab03} Habing, H.~J., \& Olofsson, H.\ 2003, Asymptotic giant branch stars, by Harm J.~Habing and Hans Olofsson.~Astronomy and astrophysics library, New York, Berlin: Springer, 2003
\bibitem[Honma et al.(2007)]{hon07} Honma, M., et al.\ 2007, \pasj, 59, 889 
\bibitem[Honma et al.(2008)]{hon08} Honma, M., Kijima, M., Suda, H., et al.\ 2008, \pasj, 60, 935 
\bibitem[Ita et al.(2004)]{ita04-1} Ita, Y., Tanab{\'e}, T., Matsunaga, N., et al.\ 2004, \mnras, 347, 720 
\bibitem[Jike et al.(2005)]{jik05}Jike, T., Fukuzaki, Y., Shibuya, K., Doi, K., Manabe, S.,Jauncey, D. L., Nicolson, G. D., \& McCulloch, P. M. 2005, Polar Geosci., 18, 26
\bibitem[Kamezaki et al.(2012)]{kam12} Kamezaki, T., Nakagawa, A., Omodaka, T., et al.\ 2012, \pasj, 64, 7 
\bibitem[Kamezaki et al.(2014)]{kam14} Kamezaki, T., Kurayama, T., Nakagawa, A., et al.\ 2014, \pasj, 118 
\bibitem[Kamezaki et al.(2015)]{kam15} Kamezaki, T., Nakagawa, A., Omodaka, T., et al.\ 2015, \pasj, 208 
\bibitem[Kamohara et al.(2010)]{kam10} Kamohara, R., Bujarrabal, V., Honma, M., et al.\ 2010, \aap, 510, A69 
\bibitem[Karovska et al.(1997)]{kar97} Karovska, M., Hack, W., Raymond, J., \& Guinan, E.\ 1997, \apjl, 482, L175 
\bibitem[Kawaguchi et al.(2000)]{kaw00} Kawaguchi, N., Sasao, T., \& Manabe, S.\ 2000, \procspie, 4015, 544
\bibitem[Knapp et al.(2000)]{kna00} Knapp, G.~R., Crosas, M., Young, K., \& Ivezi{\'c}, {\v Z}.\ 2000, \apj, 534, 324 
\bibitem[Kobayashi et al.(2003)]{kob03} Kobayashi, H., et al.\ 2003, ASP Conference Series, 306, 48P 
\bibitem[Kurayama et al.(2005)]{kur05} Kurayama, T., Sasao, T., \& Kobayashi, H.\ 2005, \apjl, 627, L49 
\bibitem[Kusuno et al.(2013)]{kus13} Kusuno, K., Asaki, Y., Imai, H., \& Oyama, T.\ 2013, \apj, 774, 107 
\bibitem[Manabe et al.(1991)]{man91}Manabe, S., Yokoyama, K., \& Sakai, S. 1991, IERS Techn. Note, 8, 61 
\bibitem[Min et al.(2014)]{min14} Min, C., Matsumoto, N., Kim, M.~K., et al.\ 2014, \pasj, 66, 38 
\bibitem[Nakagawa et al.(2008)]{nak08} Nakagawa, A., Tsushima, M., Ando, K., et al.\ 2008, \pasj, 60, 1013 
\bibitem[Nakagawa et al.(2014)]{nak14} Nakagawa, A., Omodaka, T., Handa, T., et al.\ 2014, \pasj, 66, 101 
\bibitem[Nyu et al.(2011)]{nyu11} Nyu, D., Nakagawa, A., Matsui, M., et al.\ 2011, \pasj, 63, 63 
\bibitem[Olnon(1977)]{oln77} Olnon, F.~M.\ 1977, Ph.D.~Thesis, Univ. Leiden
\bibitem[Perryman et al.(1997)]{per97} Perryman, M.~A.~C., et al.\ 1997, \aap, 323, L49 
\bibitem[Reid et al.(1977)]{rei77} Reid, M.~J., Muhleman, D.~O., Moran, J.~M., Johnston, K.~J., \& Schwartz, P.~R.\ 1977, \apj, 214, 60 
\bibitem[Sahai et al.(2008)]{sah08} Sahai, R., Findeisen, K., Gil de Paz, A., \& S{\'a}nchez Contreras, C.\ 2008, \apj, 689, 1274 
\bibitem[Shibata et al.(1998)]{shi98} Shibata, K.~M., Kameno, S., Inoue, M., \& Kobayashi, H.\ 1998, IAU Colloq.~164: Radio Emission from Galactic and Extragalactic Compact Sources, 144, 413 
\bibitem[Shintani et al.(2008)]{shi08} Shintani, M., Imai, H., Ando, K., et al.\ 2008, \pasj, 60, 1077 
\bibitem[van Leeuwen(2007)]{van07} van Leeuwen, F.\ 2007, Hipparcos, the New Reduction of the Raw Data.~By Floor van Leeuwen, Institute of Astronomy, Cambridge University, Cambridge, UK Series: Astrophysics and Space Science Library 
\bibitem[Vlemmings et al.(2003)]{vle03} Vlemmings, W.~H.~T., van Langevelde, H.~J., Diamond, P.~J., Habing, H.~J., \& Schilizzi, R.~T.\ 2003, \aap, 407, 213 
\bibitem[Vlemmings \& van Langevelde(2007)]{vle07} Vlemmings, W.~H.~T., \& van Langevelde, H.~J.\ 2007, \aap, 472, 547
\bibitem[Whitelock et al.(1994)]{whi94} Whitelock, P., Menzies, J., Feast, M., et al.\ 1994, \mnras, 267, 711 
\bibitem[Whitelock \& Feast(2000)]{whi00} Whitelock, P., \& Feast, M.\ 2000, \mnras, 319, 759 
\bibitem[Willson et al.(1981)]{wil81} Willson, L.~A., Garnavich, P., \& Mattei, J.~A.\ 1981, Information Bulletin on Variable Stars, 1961, 1 
\bibitem[Wood et al.(1999)]{woo99} Wood, P.~R., Alcock, C., Allsman, R.~A., et al.\ 1999, Asymptotic Giant Branch Stars, 191, 151 
\bibitem[Zhang et al.(2012)]{zha12} Zhang, B., Reid, M.~J., Menten, K.~M., Zheng, X.~W., \& Brunthaler, A.\ 2012, \aap, 544, AA42 
\end{thebibliography}
\end{document}